\documentclass[aps,pra,preprint,showpacs,amssymb,amsmath,superscriptaddress]{revtex4-1}

\usepackage{color}
\usepackage{graphicx}
\usepackage{times}

\newcommand{\ket}[1]{|#1\rangle}

\newcommand{\deldelsq}[2]{\frac{\partial^2#1}{\partial{#2}^2}}
\newcommand{\pd}[2]{\frac{\partial#1}{\partial#2}}
\newcommand{\conj}[1]{{#1}^\ast}
\newcommand{\rme}{{\mathrm e}}
\newcommand{\abs}[1]{\modulus{#1}}
\newcommand{\modulus}[1]{\left|#1\right|}
\newcommand{\eplus}[0]{e_+(\xi,\tau)}
\newcommand{\conjeplus}[0]{\conj{e}_+(\xi,\tau)}
\newcommand{\eminus}[0]{e_-(\xi,\tau)}
\newcommand{\conjeminus}[0]{\conj{e}_-(\xi,\tau)}
\newcommand{\wfm}[0]{\psi_m(\xi,\tau)}
\newcommand{\wfmminus}[0]{\psi_{m-2}(\xi,\tau)}
\newcommand{\wfmplus}[0]{\psi_{m+2}(\xi,\tau)}
\newcommand{\conjwfmminus}[0]{\conj{\psi}_{m-2}(\xi,\tau)}
\newcommand{\conjwfmplus}[0]{\conj{\psi}_{m+2}(\xi,\tau)}

\newcommand{\pdxi}[1]{\pd{#1}{\xi}}
\newcommand{\kl}[0]{k_l}

\newcommand{\ie}{i.e., }

\newcommand{\sr}[0]{superradiance }
\newcommand{\srnospace}[0]{superradiance}
\newcommand{\sls}[0]{superradiant light scattering }
\newcommand{\slsnospace}[0]{superradiant light scattering}
\newcommand{\MS}[0]{Maxwell-Schr\"{o}dinger }

\newcommand{\Nat}[0]{N_\text{at}}
\newcommand{\Nph}[0]{N_\text{ph}}
\newcommand{\Is}[0]{I_\text{s}} 

\newcommand{\pdt}[1]{\pd{#1}{t}}
\newcommand{\omegal}[0]{\omega_l}
\newcommand{\wfzeron}[0]{\psi_0} 
\newcommand{\wftwon}[0]{\psi_2}
\newcommand{\eminusn}[0]{e_-}
\newcommand{\eplusn}[0]{e_+}
\newcommand{\conjeplusn}[0]{\conj{e}_+}
\newcommand{\conjeminusn}[0]{\conj{e}_-}
\newcommand{\conjwfzeron}[0]{\conj{\psi}_0}
\newcommand{\conjwftwon}[0]{\conj{\psi}_2}

\newcommand{\red}[1]{\textcolor{black}{#1}}
\begin{document}



\title{Semi-classical Dynamics of Superradiant Rayleigh Scattering in a Bose-Einstein Condensate}

\author{J. H. M\"uller}
\affiliation{Niels Bohr Institute, Blegdamsvej 17, 2100 Copenhagen {\O}, Denmark}
\author{D. Witthaut}
\affiliation{Forschungszentrum J\"ulich, Institute for Energy and Climate Research – Systems Analysis and Technology Evaluation (IEK-STE), 52428 J\"ulich, Germany}
\affiliation{Institute for Theoretical Physics, University of Cologne, 50937 K\"oln, Germany}
\author{R. le Targat}
\affiliation{LNE-SYRTE, Observatoire de Paris, CNRS, UPMC, LNE, 61 avenue de l'Observatoire 75014 Paris, France}
\author{J. J. Arlt}
\affiliation{Institute for Physics and Astronomy, Ny Munkegade 120, 8000 Aarhus C, Denmark}

\author{E. S. Polzik}
\affiliation{Niels Bohr Institute, Blegdamsvej 17, 2100 Copenhagen {\O}, Denmark}
\author{A. J. Hilliard}\email[]{hilliard@phys.au.dk}
\affiliation{Niels Bohr Institute, Blegdamsvej 17, 2100 Copenhagen {\O}, Denmark}
\affiliation{Institute for Physics and Astronomy, Ny Munkegade 120, 8000 Aarhus C, Denmark}


\begin{abstract}
 Due to its coherence properties and high optical depth, a Bose-Einstein condensate provides an ideal setting to investigate collective atom-light interactions. Superradiant light scattering in a Bose-Einstein condensate is a fascinating example of such an interaction.
It is an analogous process to Dicke superradiance, in which  an electronically inverted sample decays collectively, leading to the emission of one or more light pulses in a well-defined direction. 
Through time-resolved measurements of the superradiant light pulses emitted by an end-pumped BEC, we \red{study} the close connection \red{of \sls} with Dicke superradiance. A  1D model of the system yields good agreement with the experimental data and  shows that the dynamics results from the structures that build up in the light and matter-wave fields along the BEC. This paves the way for exploiting the atom-photon correlations generated by the superradiance. 
\end{abstract}

\maketitle
\section{Introduction}
\begin{figure}[t]
 \centering
   \includegraphics[width=7.7cm]{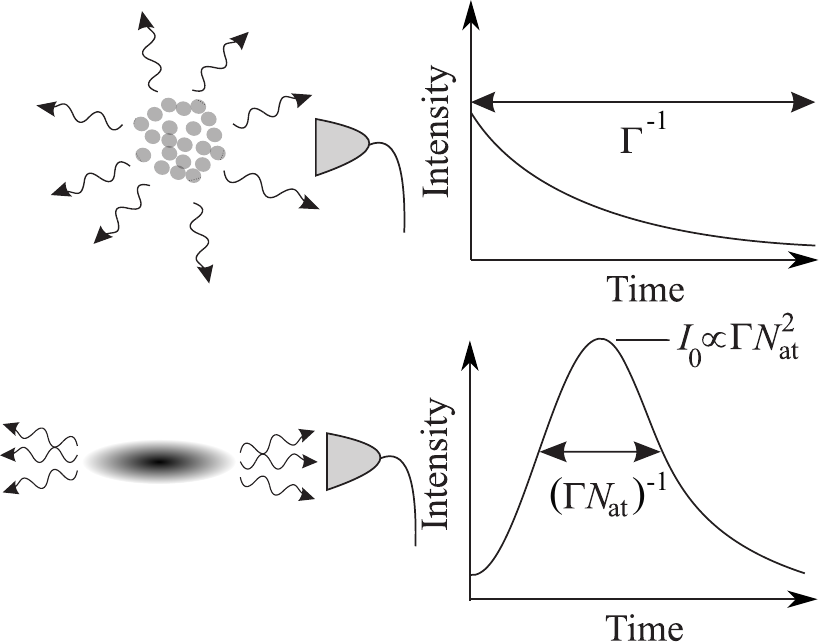}\\
  \caption{Illustration of the differences between single particle and collective spontaneous emission. Upper figure: a dilute sample of atoms in the electronic excited state decay independently over a timescale $\Gamma^{-1}$. Lower figure: in a dense, anisotropic sample, the light  is  emitted as one or more pulses of light  along the direction of highest optical depth. The pulses have peak intensity $I_0\propto\Gamma\Nat^2$ and a characteristic width $\propto (\Gamma\Nat)^{-1}$. Adapted from `Superradiance: An essay in the theory of collective spontaneous emission', Ref. \cite{SR_review_GrossHaroche} with permission from Elsevier.}
  \label{fig:SRillust}
\end{figure}

Superradiant light scattering in a Bose Einstein condensate provides a striking example of collective enhancement  in the interaction of light and matter in ultracold atomic samples \cite{S.Inouye07231999,DominikSchneble04182003}.
The phenomenon is analogous to the collective spontaneous emission studied by R. H. Dicke in his seminal paper \cite{PhysRev.93.99}. In single atom spontaneous emission,  the intensity of emitted light falls off exponentially  at the natural decay rate $\Gamma$. In contrast, 
 a dense ensemble of $\Nat$ atoms in the electronic excited state can collectively relax to the ground state through the emission of one or more pulses of light  \cite{SR_review_GrossHaroche}. In `Dicke superradiance' -- the multi-atom generalization of the Wigner-Weisskopf approach to spontaneous emission -- the emitted light pulses have an amplitude that scales as $\Gamma\Nat^2$, and a characteristic width  $\propto(\Gamma\Nat)^{-1}$ (see Figure \ref{fig:SRillust}). 

In superradiant light scattering (SLS), the electronically inverted sample is replaced by an atomic ensemble `dressed' by a pump beam. The pump induces spontaneous scattering in the sample, populating one or more previously unoccupied modes of light. 
 Thus, the spontaneous scattering rate $R$ takes the role of   $\Gamma$ in Dicke superradiance, and since $R\ll\Gamma$ for standard experimental parameters,  the dynamics of SLS is in general much slower than Dicke superradiance. 
 Stimulated Raman scattering from the pump beam into the spontaneously populated mode builds up over a timescale $\propto R^{-1}$. 
In superradiant light scattering, the  recoil momentum an atom gains from photon scattering plays a crucial role in the dynamics. 
In this sense, the narrow momentum spread and long-lived motional coherence in a  BEC is ideal for maintaining the coherence between atomic clouds with different momenta. The dipole emission pattern of the driven optical transition and the sample anisotropy determine which spatial light modes experience the most gain.

In the standard realization of SLS, condensates are produced in cylindrically symmetric  harmonic potentials, leading to cigar-shaped clouds where the dominant superradiant light modes are those that propagate along the long axis of the BEC - the so-called endfire modes. In the superradiant Rayleigh scattering process, an atom gains a recoil kick from  scattering a photon and returns to its initial internal state. 
 In an end-pumped sample, where the pump beam propagates co-linear with the endfire modes, the interference of the  stationary condensate and a recoiling order leads to a modulation of the atomic density with spatial period $\lambda/2$, where $\lambda$ is the wavelength of the incident light. Similarly, the backscattered light interferes with the pump beam, leading to an intensity modulation, again with spatial period  $\lambda/2$. In this way, the dynamics is determined by the overlap of these light and matter-wave `gratings' \cite{DominikSchneble04182003}, which vary in amplitude and phase over the length of the BEC.
 
From this brief discussion, it is clear that SLS is a multi-mode process and that the emission of one or more light pulses from an extended atomic sample is the result of complex dynamics. Experimentally, when imaging the backscattered light on a camera, we observe a  unimodal intensity distribution for a given experimental run, but the position and shape of this distribution varies slightly from run-to-run.
Without resorting to complicated theoretical models  that do not make an \textit{a priori} single mode assumption \cite{PhysRevA.68.023809,PhysRevA.80.033804}, a  standard approach to study Dicke superradiance or \sls in extended samples is to consider an initial `quantum' phase followed by a one-dimensional `semi-classical' evolution whereby an assumed single mode of light is amplified \cite{PhysRevA.24.1980}. 
Our experimental results are  well modeled  by such an approach, achieving good agreement over a wide parameter range. 

In this article, we focus on the semi-classical dynamics associated with the amplification and propagation of light in the condensate. This article extends our previous work \cite{hilliard:051403} by focusing on the close connection between Dicke superradiance and SLS and provides a detailed comparison of simulations with experiment.
The paper is structured as follows.  In Section \ref{sect:exptal_setup_for_SLS}, we describe the experimental setup. In Section \ref{sect:model}, we present the 1D \MS equations used to model the experiment. In Section \ref{Coupled}, we present experimental results, illustrating the strong analogy between Dicke \sr and \slsnospace. In particular, we show that the measured  superradiant pulses demonstrate the same scaling with scattering rate and atom number as in Dicke \srnospace. In Section \ref{sect:spatial_dep_dynamics}, we use the 1D model to illustrate the spatially dependent dynamics within the BEC that lead to the observed light scattering. Section \ref{conclusion} offers conclusions and an outlook for future work.
 
\section{Experimental set-up and parameters}\label{sect:exptal_setup_for_SLS}

\begin{figure}
 \centering
  \includegraphics[width=7.7cm]{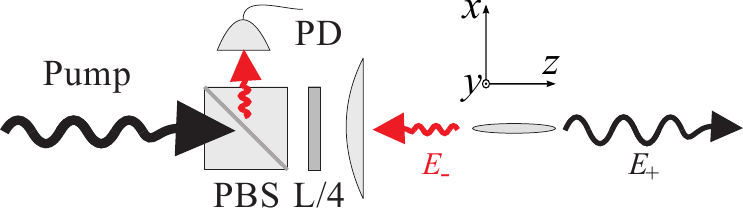}\\
  \caption{Schematic of the experimental setup.}
  \label{fig:setup}
\end{figure}

We initiate  superradiant Rayleigh scattering in a trapped BEC by exposing it to a pulse of off-resonant light propagating along the long axis of the condensate. Figure \ref{fig:setup} shows a schematic of the key features of the experiment. The BEC is generated in a Ioffe-Pritchard magnetic trap through  evaporatively cooling a cloud of $^{\textrm{\small{87}}}$Rb atoms in the $\ket{F=1,m_F=-1}\equiv\ket{1,-1}$  hyperfine state; the bias field is oriented along the long axis ($z$) of the trap. Typically, we use  condensates containing $\sim 1.35\times10^6$ atoms, with in-trap Thomas Fermi radii of \mbox{$\rho_0=6.4~\mu$m} and \mbox{$z_0=65$ $\mu$m} in the radial and axial directions, respectively, where there is no discernible thermal fraction.  The pump light is detuned by  $\delta=\omega-\omegal$ from the $\ket{1,-1}\rightarrow \ket{2,-2}'$ transition on the $D_1$ line of $^{87}$Rb at 795~nm, with $\omega$ the atomic transition frequency and $\omegal$ the  laser frequency; the  pump light is circularly polarized with respect to the magnetic bias field. All data presented is for $\delta=-2\pi\times2.6$~GHz using rectangular pump pulse envelopes. The superradiant dynamics occur in-trap, with the trapping potential extinguished immediately after the end of the pump pulse. The pump beam has a Gaussian intensity profile and is focused to a waist of 
13.2~$\mu$m at the centre of the condensate with a negligible change in beam waist over the length of the cloud.  Light is backscattered  by the sample in the same polarization as the input beam; it is thus reflected by the polarizing beamsplitter (PBS) and then impinges on a sensitive detector. The detector has a bandwidth of 400~kHz and is shot-noise limited for photon fluxes greater than $10^5$ photons/$\mu$s; the bandwidth leads to the smoothing of the detected light over a timescale of $\sim 2.5~\mu$s. 
Pictures of the atoms are obtained after 45ms time-of-flight by resonant absorption imaging.

This choice of transition and detuning approximates  well a two-level atom driven  by an applied laser field. The single particle spontaneous scattering rate for a two-level atom is given by \mbox{$R= \Gamma{I}[({2\Is})({1+I/\Is + (2\delta /\Gamma)^2})]^{-1}$,} where $I$ is the intensity of the applied laser field and  $\Is$ is the saturation intensity of the optical transition;
for the studied experimental parameters, $R$  lies in the  range $0.5 - 3.2\times10^3$~s$^{-1}$. An atom in the excited state, $\ket{2,-2}'$, can decay to three levels: $\ket{2,-2}$, $\ket{2,-1}$, and $\ket{1,-1}$, with probabilities $1/3$, $1/6$, and $1/2$, respectively. The small Clebsch Gordon coefficients for decay into the $F=2$ manifold prevent significant populations accumulating in $\ket{2,-2}$ and $\ket{2,-1}$ via  superradiant  Raman scattering  \cite{yoshikawa:041603}.
Given the low rate of real excitations $R$ on the experimental timescale ($\sim 10-100~\mu$s), and the dephasing that occurs between these states in the magnetic trap \cite{PhysRevLett.108.090401},  we  neglect these states in our modeling.

\section{1D model of the dynamics}\label{sect:model}

In the end-pumped geometry, SLS occurs primarily in  the light that is backscattered by the atomic ensemble; this leads to the concomitant scattering of atoms into forward momentum orders separated by $2\hbar\kl$, where $\kl$ is the wavenumber of the pump light. Superradiant backscattering predominates because there is no net momentum change for scattering in the forward direction.
An atom with momentum $2\hbar\kl$ is generated by the destruction of a pump photon $E_+$ and the creation of a backscattered photon $E_-$. In Rayleigh scattering,  the process can occur repeatedly, and, with appropriate parameters, one can generate several  diffracted orders in the forward direction.
For single particle scattering rates  much smaller than the recoil frequency $\omega_r=2\pi\times$3.6~kHz, scattering to higher atomic momentum orders occurs sequentially on a time scale \mbox{$\sim \tau_r=2\pi/\omega_r\sim100~\mu$s}. When $R\sim\omega_r$, there is sufficient gain  for atoms to be backscattered  into negative momentum orders, i.e., the Kapitza-Dirac regime where atoms absorb backscattered light and re-emit into the forward direction.

The Fresnel number of the experimentally studied condensates is approximately one, implying that the main aspects of the system's dynamics may be described by a one dimensional theory \cite{PhysRevA.20.2047}. 
The Fresnel number is given by $F=\pi w^2/(\lambda L)$
where $w$ is the radius and $L$ the length  of the (assumed cylindrical) atomic sample; alternatively, $w$ gives the radius of an aperture or waist of a Gaussian beam and $L$ the distance to the plane of observation. If $F\ll 1$, the scattered light is confined to a narrow cone in the forward direction but this comes with the complication of introducing a strong radial dependence on the transverse modes ($F\ll 1$ marks the region of applicability of Fraunhofer diffraction). Alternatively, if $F>1$, the axial modes have little radial dependence but non-axial modes are supported.
For $F\approx1$, the light scattered within the sample retains its  transverse distribution along the length of the sample, and furthermore  this distribution has little transverse variation; \ie the light within the sample is well-described by a single, approximately flat, transverse mode.

\subsection{Maxwell-Schr\"{o}dinger equations}\label{sect:Maxwell_Sdgr_eqns}
The starting point for the analysis of this system is the  Schr\"odinger equation for the ground state of a two-level atom in the presence of an off-resonant light field, and the wave equation with a polarization source term \cite{zobay:041604,PhysRevA.73.013620}:
\begin{equation}\label{eqn:SR}
i\hbar\frac{\partial\psi}{\partial t} =  -\frac{1}{2M}\nabla^2{\psi}
+\frac{1}{\hbar\delta}(\mathbf{d\cdot E^{(-)}})(\mathbf{d\cdot E^{(+)}})\psi,
\end{equation}
\begin{equation}\label{eqn:wave}
c^2\nabla^2\mathbf{E^{(\pm)}}-\deldelsq{\mathbf{E^{(\pm)}}}{t}=
\frac{1}{\varepsilon_0}\deldelsq{\mathbf{P^{(\pm)}}}{t},
\end{equation}
where the total electric field is given by $\mathbf{E}=\mathbf{E^{(+)}}+\mathbf{E^{(-)}}$, $\mathbf{d}$ is the atomic dipole moment, $M$ the mass, and the polarization is given by
\begin{equation}\label{eqn:pol_term}
    \mathbf{P^{(+)}}(\mathbf{r},t)=-\frac{\mathbf{d}\abs{\psi(\mathbf{r},t)}^2\mathbf{d\cdot E}^{(+)}(\mathbf{r},t)}{\hbar\delta},\quad\mathbf{P}^{(-)}=\conj{\mathbf{P}^{(+)}}.
\end{equation}
The excited electronic state has been  adiabatically eliminated given the assumed low rate of real excitations.
These equations give a self-consistent description of an ensemble of two-level atoms interacting with a classical electric field. The applied field polarizes the atoms according to quantum mechanics, the dipole moments of these atoms are summed to give the macroscopic polarization $\mathbf{P}(\mathbf{r},t)$, and this enters the wave equation as a source term \cite{PhysRev.185.517}. We assume that the polarization of the atoms is linear in the applied electric field, and therefore that there is no saturation of the atomic transition - a point which needs to be confirmed as a matter of self-consistency in the solution of the problem. Indeed, for the parameters considered in this work, this condition is always fulfilled. At this stage, we neglect the harmonic trapping potential and the mean field  term representing interactions between atoms.

To solve the above equations, we  make the following approximations. Based on the discussion of Section~\ref{sect:exptal_setup_for_SLS}, for an atomic sample with Fresnel number $F\sim 1$, it is reasonable to ignore the transverse spatial variation of the condensate wavefunctions and electric fields. As such, we make the ansatz
\begin{equation}\label{eqn:wfm}
\psi(z,t)=\sum_{m=2n}\psi_m(z,t)\rme^{-i(\omega_m t-m k_l z)},
\end{equation}
and
\begin{equation}\label{eqn:E}
\mathbf{E}^{(+)}(z,t) = \left[E_+(z,t)\rme^{-i(\omegal t- k_l z)}
                      + E_-(z,t)\rme^{-i(\omegal t+k_l z)}\right].
\end{equation}
$\psi_m(z,t)$ is the \emph{slowly varying amplitude}  for the atomic momentum order \mbox{$m=2n$} for integer $n$; the concomitant recoil frequency is given by \mbox{$\omega_m=m^2\omega_r$}, with $\omega_r=\hbar\kl^2/(2M)$.  \mbox{$E_+$ and $E_-$} are slowly varying amplitudes for the forward and backward travelling electric fields; the  electric field polarization has  been suppressed in Equation \eqref{eqn:E}. 
Note that in this formalism, we cannot distinguish between incident and forward scattered light: $E_+$ contains both components. Such an identification is necessary in a 1D treatment, which assumes that the incident and scattered fields occupy the same light mode. This identification is supported somewhat by the fact that the pump is partially mode matched to the BEC (see Section~\ref{sect:exptal_setup_for_SLS}); we take account of the spatial overlap between the BEC and the pump intensity distribution when calculating the boundary conditions for $E_+$. 

Upon substitution of Equation  (\ref{eqn:wfm}) into (\ref{eqn:SR}), we identify terms that oscillate at $(\omega_m t-m k_l z)$. Similarly we substitute (\ref{eqn:E}) into (\ref{eqn:wave}) and identify terms with the common phase $(\omegal t\pm\kl z)$. For the light fields, we make the `Slowly Varying Envelope Approximation'. 
 The content of the approximation is to neglect derivatives of the slowly varying envelopes $E_+$ and $E_-$
with respect to terms involving the derivatives of the fast oscillating exponentials. That is, we assume
\begin{align}\label{eqn:svea_def}
    \abs{\pdt{E_\pm}}& \ll\abs{\omegal E_\pm} \hspace{2cm}\textrm{and}& \abs{\pd{E_\pm}{z}}& \ll\abs{\kl E_\pm}.
\end{align}
To simplify the ensuing equations, we  rescale the position and time variables such that $\xi=k_l z$ and $\tau=2\omega_r t$. The light field amplitudes are rescaled according to: $ e_\pm = E_\pm[\hbar\omegal k_l/(2\varepsilon_0 A)]^{-\frac{1}{2}}$,
with $A$ the  cross-sectional area of the (assumed cylindrical) BEC. Finally, we obtain:
\begin{equation}\label{eqn:SR_wf}
\begin{split}
i\frac{\partial\wfm}{\partial\tau} = & -\frac{1}{2}\deldelsq{\psi_m(\xi,\tau)}{\xi}
-i m \pd{\wfm}{\xi}\\
 & +\Lambda \conjeminus\eplus\wfmminus\rme^{+2i(m-1)\tau}\\
 & +\Lambda\conjeplus\eminus\wfmplus\rme^{-2i(m+1)\tau}\\
 & +\Lambda(\abs{\eplus}^2+\abs{\eminus}^2)\wfm ,
\end{split}
\end{equation}

\begin{align}
\pdxi{\eplus} =  -i\frac{\Lambda}{\chi}  \sum_{m=2n} & \eminus\wfm\conjwfmminus\rme^{-2i(m-1)\tau}\notag\\
                                + & \eplus\abs{\wfm}^2,\label{eqn:eplus}\\
\pdxi{\eminus} =  +i\frac{\Lambda}{\chi}\sum_{m=2n}&
\eplus\wfm\conjwfmplus\rme^{+2i(m+1)\tau}\notag\\
                                  + & \eminus\abs{\wfm}^2,\label{eqn:eminus}
\end{align}
with the coupling constants $\Lambda=\abs{\mathbf{d}}^2\omegal k_l/(4\omega_r\hbar\delta\varepsilon_0 A)$ and $\chi=ck_l/(2\omega_r)$. Retardation effects have been neglected in Equations (\ref{eqn:eplus}) and (\ref{eqn:eminus}) given the length of the condensate $L=130~\mu$m, which allows us to discard a time derivative term; however,  with the definition of a retarded time (in unscaled quantities), $t'=t-z/c$, the result can be made exact \cite{SR_review_GrossHaroche}.

\subsection{Four-wave mixing}\label{sect:four_wave_mixing}
Equations (\ref{eqn:SR_wf})--(\ref{eqn:eminus}) describe a Raman interaction where a ladder of momentum states is coupled by two counter-propagating light fields. The first  term in Equation~(\ref{eqn:SR_wf}) describes the quantum diffusion of the wavefunction envelope and the second term  is the momentum displacement of the wavefunction  due to the atomic recoil. In our parameter regime and interaction time,  these envelopes are slowly varying and thus  contribute very little to the dynamics. Terms three and four describe the coupling between neighbouring momentum states via the exchange of photons between $e_+$ and $e_-$. The final terms in Equation~(\ref{eqn:SR_wf}) account for phase rotation of the matter wave due to the light
shift.
Equations (\ref{eqn:eplus}) and (\ref{eqn:eminus}) have terms equivalent to the coupling terms in Equation (\ref{eqn:SR_wf}). Specifically, the creation  of photons in $e_-$ and recoiling atoms in $\psi_{m+2}$,
and the annihilation of $e_+$ photons and $\psi_m$ atoms. The last
terms in  Equations (\ref{eqn:eplus}) and (\ref{eqn:eminus})
describe the effect   of the slowly varying refractive index on light due
to the large scale atomic density distribution.
If one disregards the quantum diffusion and the momentum displacement terms in Equation (\ref{eqn:SR_wf}), these equations have the symmetry of a four-wave mixing process.

Such a four-wave mixing process is  non-linear: here, the scattering of atoms depends on the local intensity of light and the scattering of photons depends on the local atomic density. In general, a standing wave of light will arise with spatial period $\lambda/2$, but comprised of several frequency components shifted by multiples of $4\omega_r$, which are represented by the exponential terms $\exp[{\pm2i(m\pm1)\tau}]$ in Equations (\ref{eqn:eplus}) and (\ref{eqn:eminus}).
Given that the effect  of these frequency changes on the wavelength is insignificant over the spatial extent of the condensate, they will manifest themselves as time-varying amplitude and phase modulations of the standing wave along the length of the sample, i.e., a `walking standing wave' with a dynamically evolving amplitude. A similar picture arises on the atomic side:  since we consider Rayleigh scattering where the internal state of the atoms remains unchanged, condensates with different momenta interfere,   leading to a density modulation.
  In general, this matter-wave grating is comprised of as many spatial periods and oscillating frequencies as there are populated momentum orders. In the regime of weak excitation, corresponding to a low single particle scattering rate $R$, only $\psi_0$ and $\psi_2$ become significantly populated leading to a spatial period of $\lambda/2$. This leads to the physical picture that  the atomic density modulation corresponds to a Bragg grating with a sinusoidal modulation of the refractive index.
\begin{figure}[t]
 \centering
  \includegraphics[width=7.7cm]{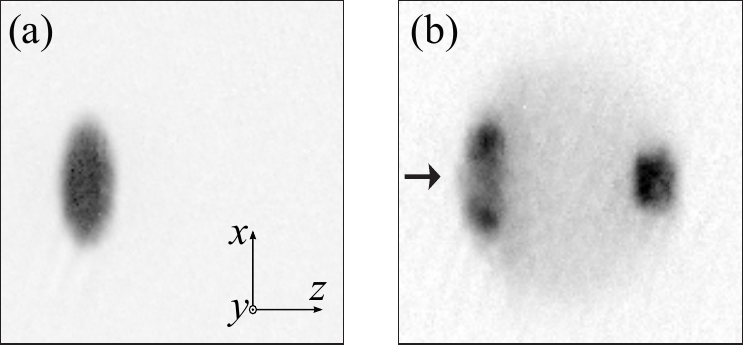}\\
  \caption{Illustration of  the Bragg condition in superradiant scattering for weak pumping ($R\ll\omega_r$). A pump beam that is much broader than the transverse extent of the BEC is flashed on the magnetically trapped atoms; the trapping potential is extinguished immediately after the interaction. (a) An unperturbed BEC, (b) a BEC that has been dressed by the pump beam (indicated by the black arrow); both pictures are  absorption images taken after 45 ms time-of-flight. With the pump well-aligned, the diffraction to $2\hbar\kl$ occurs primarily for atoms centred around zero transverse momentum.}
  \label{fig:nice_sr_pic}
\end{figure}

This picture of Bragg scattering is illustrated in Figure \ref{fig:nice_sr_pic}. Figure \ref{fig:nice_sr_pic}(a) shows a time-of-flight absorption image of a representative  BEC used in our experiments; at this long time-of-flight, the absorption image reflects the in-trap momentum distribution.  Figure \ref{fig:nice_sr_pic}(b) shows a time-of-flight image of a BEC that has been end-pumped by a weak probe beam, such that only one forward momentum order has been populated by \slsnospace. In this case, the pump beam has been  collimated to a waist much broader than the BEC and aligned along the long axis of the trap. It is evident that it is primarily atoms centred around zero transverse momentum that are diffracted into the $2\hbar\kl$ momentum state. 
For incident plane waves, the Bragg condition for light to be backscattered is fulfilled for those atoms with close to zero transverse momentum. Collisions between 0 and $2\hbar\kl$ atoms, and to a lesser extent the nearly isotropic spontaneous Rayleigh scattering, lead to the visible isotropic scattering halo. 
Figure \ref{fig:nice_sr_pic}(b) illustrates the strengths and the weaknesses of the 1D model: the \MS equations provide a straightforward and physically intuitive picture of the dynamics but the 1D treatment is nonetheless an approximate description.

\subsection{Simulations}\label{Simulations}
We solve Equations  (\ref{eqn:SR_wf})--(\ref{eqn:eminus}) numerically for experimental parameters by alternatingly updating the atomic wave functions $\psi_m(\xi)$ and the light field amplitudes $e_{\pm}(\xi)$. Given a solution for the light field at time $\tau$, we propagate the atomic wave functions over a short time step $d\tau$ using a split operator technique  \cite{Feit_spectral}:
\begin{equation}
\begin{split}
	                 	                     U(\tau+d\tau,\tau) & = \exp\left(- \frac{i}{\hbar} \int_\tau^{\tau+d\tau} T + V(\tau') d\tau'\right),\\
	                 & = \exp\left(-\frac{ iTd\tau}{2\hbar}\right) \exp\left(-\frac{i}{\hbar} \int_\tau^{\tau+d\tau} V(\tau') d\tau' \right)\exp\left(-\frac{ iTd\tau}{2\hbar}\right) + O(d\tau^3),
\end{split}
\end{equation}
where $T$ is the time-independent kinetic energy operator and $V$ is the time-dependent coupling to the light field, and when included, the trapping potential and mean-field interaction. $V$ is diagonal in real space, so that it can be directly applied to the real space wave function. In contrast, $T$ is diagonal in momentum space so that one can apply it to the momentum space wave function and flip between the two representations using the Fast Fourier Transform. For the moderate number of discrete grid points used here, however, it is also possible to evaluate $\exp
[-i Td\tau/(2\hbar)]$ numerically in real space. The initial wavefunction $\psi_0$ is taken to be a 1D Thomas-Fermi profile normalized to the number of atoms in the trap $N_\text{at}$. The number of required momentum orders  depends on the strength and duration of the interaction and is chosen so that the outermost orders are negligibly populated.

Following the time evolution step of the atomic wavefunctions, the light field amplitudes are  updated by solving the ordinary differential equations (\ref{eqn:eplus}) and (\ref{eqn:eminus})  at time $\tau+d\tau$ with the boundary conditions $e_+(0,\tau)=e_{i}$ and $e_-(k_lL,\tau)=0$. 
$e_{i}$ is a constant derived from the experimental pump photon flux and the (assumed) geometrical overlap of the BEC and the Gaussian intensity distribution of the pump beam, which models the fraction of photons in the pump beam that can be regarded as occupying mode $e_+$. The backscattered photon flux is 
\begin{equation}\label{eqn:Nph}
\Nph(\tau) =C |e_-(0,\tau)|^2 ,
\end{equation}
with $C$ a constant.  As a check of self-consistency of the numerical implementation of the \MS equations, we have verified that the total photon flux is conserved: {$ \abs{e_+(0,\tau)}^2=\abs{e_-(0,\tau)}^2+\abs{e_+(\kl L,\tau)}^2$.}

 Equations (\ref{eqn:SR_wf})--(\ref{eqn:eminus})  contain no  noise term, so to instigate the process in this semi-classical approach we seed the dynamics by taking a non-zero first order momentum component $\psi_{2}(\xi,0)=\psi_0/\sqrt{N_\text{at}}$, corresponding to a single delocalized  atom  in the first side-mode \cite{PhysRevA.73.013620}. 
With this choice of seed, we scale the field amplitudes $e_{i}$ derived from experimental parameters by a global factor of 10.5\% to 
achieve the best possible agreement in the arrival time and amplitude of the first superradiant pulse over a wide range of detunings, atom numbers and single particle scattering rates. This scaling is compatible with calibration uncertainties in the experimental parameters. For a set of standard experimental parameters, a complete simulation takes on the order of a minute.

\section{Coupled wave dynamics in superradiant light scattering}\label{Coupled}
We now present experimental results that illustrate the semi-classical evolution of \sls in a Bose-Einstein condensate. Whereas most earlier experimental studies of \sls have   used  time-of-flight images of the atomic density distribution, we study the process primarily through the time-resolved detection of superradiant pulses emitted by the sample \cite{S.Inouye07231999,sadler:110401,DominikSchneble04182003,PhysRevA.69.041601,yoshikawa:041603,Li_PhysLettA}.

\subsection{Comparison  of data and simulations}\label{sect:forms_data_sims}
\begin{figure}[t]
 \centering
   \includegraphics[width=15cm]{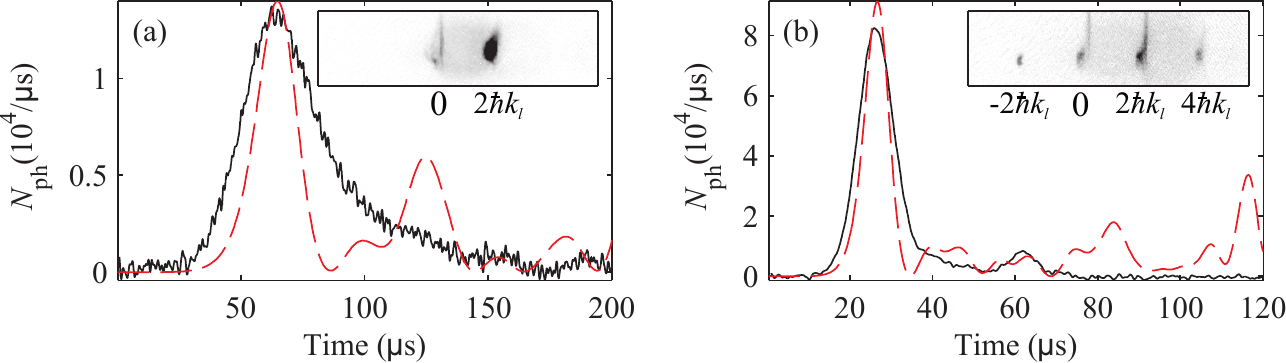}\\
  \caption{Comparison of simulations and experiment. (a) (Black solid line) Time trace of the backscattered photon flux for a  representative experimental run with $R= 0.447\times10^{3}$~s$^{-1}$; (red dashed line) simulation for the same parameters. (b) Same as (a) but for $R= 2.15\times10^{3}$~s$^{-1}$. Insets show the corresponding time-of-flight atomic density distributions following the dynamics in each time trace; the greyscale is different for the two  images  to ensure clarity. The simulations describe the arrival time and amplitude of the first superradiant pulse well, but exhibit more ringing behaviour than the experimental data.}
  \label{fig:data_forms}
\end{figure}

Figure \ref{fig:data_forms} shows representative time traces of the backscattered light, the  corresponding results from  simulations, and time-of-flight images of the atomic density following the superradiance dynamics. In the low scattering rate case, $R= 0.447\times10^{3}$~s$^{-1}$, shown in Figure \ref{fig:data_forms}(a), the superradiant dynamics leads to the transfer of population from the zero momentum class to $2\hbar\kl$. In contrast to Figure \ref{fig:nice_sr_pic}, the density distributions are complicated by a small angle between the pump beam propagation direction and  the long axis of the condensate; this angle was chosen to avoid seeding the dynamics by light scattered from the vacuum chamber. In the  time trace, the dynamics leads to the emission of a  pulse peaking at 65~$\mu$s, followed by a slow decay over about 100~$\mu$s; during the slow decay, the photon flux undergoes two or three small oscillations. The simulation captures the arrival time and the height of the superradiant pulse well, but it exhibits considerably more oscillations. 

 The situation is similar in Figure \ref{fig:data_forms}(b), which shows results for a higher single particle scattering rate, \mbox{$R=2.1\times10^{3}$~s$^{-1}$}. This scattering rate is sufficient to reach the Kapitza-Dirac regime, where atoms can absorb backscattered photons and re-emit into the pump beam, leading to atoms scattered into one or more negative momentum orders. In this case,  SLS  leads to population in one negative ($-2\hbar\kl$) and  two forward ($2\hbar\kl$ and $4\hbar\kl$) momentum orders.
 Again, the simulations describe the first superradiant pulse well, but show more oscillations than  observed in the experiment.
 
The occurence of these oscillations -- or  `ringing' -- is a general feature of superradiance in extended samples \cite{PhysRevA.14.1169}; it  arises from the fact that dynamics in one part of the sample can be driven  by light scattered from another part of the sample.
The appearance of more ringing behaviour in simulations than in experiments is a general feature of 1D models of superradiance \cite{Schuurmanns_superfluorescence}. The inclusion of the trapping potential and the mean field interaction between atoms in the \MS equations leads to  small modifications in the simulated dynamics for our experimental parameters, suggesting that it is the 1D character of the model that is the determining factor. 
 This is explored in more detail in Section~\ref{sect:spatial_dep_dynamics}.

\subsection{Dependence on single particle scattering rate}\label{sect:scrate}

\begin{figure}[t]
 \centering
  \includegraphics[width=15cm]{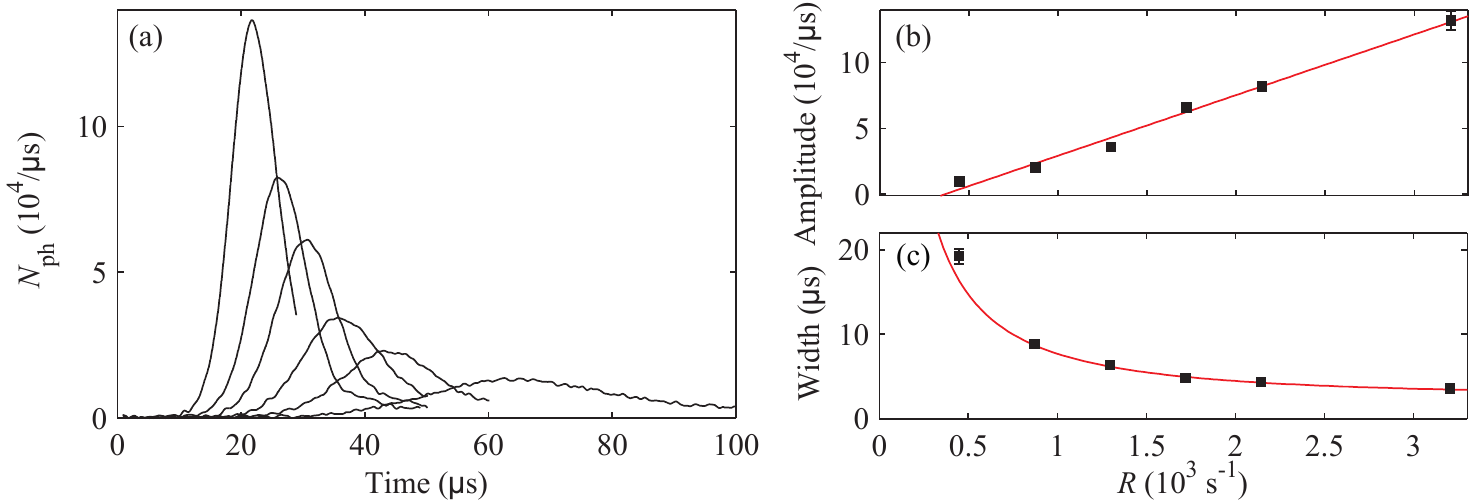}\\

    \caption{Characteristics of the first superradiant pulse as a function of $R$. (a) Representative time traces of the backscattered photon flux; in ascending order of the superradiant pulse amplitude, $R= 0.45,      0.87,    1.3,    1.7,    2.2,   3.2 \times 10^3$s$^{-1}$.
 For clarity, the traces have been clipped after the first superradiant pulse. (b)  Superradiant pulse amplitude  as a function of $R$, (red line) linear fit to the data. (c) Superradiant pulse width  as a function of $R$, (red line) $1/R$ fit to the data. Errorbars show the standard error of the mean for five realizations.}  \label{fig:pump_dep}
\end{figure}
In Dicke superradiance, 
 the amplitude of the first superradiant pulse scales with the spontaneous decay rate $\Gamma$ and the width of the pulse scales as $\Gamma^{-1}$ (see  Figure \ref{fig:SRillust}). To test this scaling experimentally, one thus requires several atomic samples with different values of $\Gamma$. In \sls in a BEC,  the single particle scattering rate $R$ has an equivalent role to $\Gamma$, but since $R$ is a known function of the pump intensity $I$ and detuning $\delta$, one can verify the scaling with the single particle scattering rate by scanning one or both of these parameters. In earlier work, we tested this analogy of $\Gamma$ with $R$ by 
 investigating the effect on the dynamics of depletion of the pump beam for the case of low $I$ and $\delta$ \cite{hilliard:051403}. In this section, we verify the scaling of the first superradiant pulse in the high detuning regime ($\delta=-2\pi\times2.6$ GHz) where depletion of the pump beam is negligible. 

Figure \ref{fig:pump_dep}(a) shows representative experimental time traces for several  values of the single particle scattering rate. 
The data shows the general trend that higher  pump power leads to  an earlier arrival of the pulses  and more sharply peaked time traces.
To quantify this,  each time trace has been fitted with a Gaussian function  $A\exp[-(t-\mu)^2/(2\sigma^2)]$. 
  Figure \ref{fig:pump_dep}(b) shows the superradiant pulse amplitude $A$ as a function of $R$. The data points are the mean of five realizations for each value of $R$.   
 As expected, the amplitude data in  Figure \ref{fig:pump_dep}(b) is well described by a  linear function in $R$. It is also evident that a threshold value of the single particle scattering rate is required to overcome  damping mechanisms in the system. These mechanisms include spontaneous emission and incoherent collisions between ground state atoms \cite{PhysRevLett.108.090401}, which lead to decay of the light and matter-wave gratings \cite{S.Inouye07231999}.
 Figure \ref{fig:pump_dep}(c) shows the superradiant pulse width $\sigma$ as a function of $R$, which has been fitted by  $ [(p_1/R)^2+p_2^2]^{\frac{1}{2}}$, with $p_1$ and $p_2$ fitted parameters.  $p_2$ represents the minimum detectable pulse width due to the detector bandwidth; its fitted value was $p_2=2.6\pm0.1~\mu$s,  in good agreement with the measured detector bandwidth (see Section \ref{sect:exptal_setup_for_SLS}).
Finally, the pulse width data exhibits the expected $R^{-1}$ scaling from Dicke superradiance.

\subsection{Dependence on atom number}\label{sect:dep_atom_number}

 The analogy of \sls with Dicke \sr is further explored by studying the scaling of the amplitude of the first superradiant pulse with atom number. As illustrated in Figure \ref{fig:SRillust},  the amplitude of the superradiant pulse in Dicke superradiance scales with $\Nat^2$.

To achieve a range of atom numbers, we hold the atoms in the magnetic trap for a variable duration  in the presence of radio-frequency (RF) radiation at the  final cut frequency used in  forced evaporative cooling. For the condensates we generate, three-body loss is the dominant loss mechanism, and leads to a fast decay of atoms (on the order of one second) in the presence of the RF radiation. The three-body loss thus helps to remove the technical run-to-run fluctuations  in the condensate number, which are on the order of a few percent. Furthermore, three-body loss is a desirable loss mechanism to employ since it generates an atom number distribution with asymptotic width $3\sqrt{\Nat}/5$, whereas single particle loss leads to a Poissonian number distribution (i.e., with mean and width $\sqrt{\Nat}$). This has recently been used to generate sub-Poissonian atom number fluctuations in  small BECs containing $\sim 100$ atoms \cite{PhysRevLett.104.120402,Itah2010}.

However, varying the atom number in a BEC changes its size and therefore the coupling strength to the pump beam. This is in contrast to experiments in Dicke superradiance, where the effective number of atoms participating in the process may be varied by the degree of population inversion, without changing the sample geometry \cite{PhysRevLett.30.309}. We take this change in coupling strength into account by calculating the in-trap size of the condensate  from the measured  number of atoms in the experiment; these dimensions are used in the BEC and light field normalizations, and in evaluating the effective number of photons interacting with the BEC from its overlap with the pump beam.
\begin{figure}[t]
\centering
  \includegraphics[width=7.7cm]{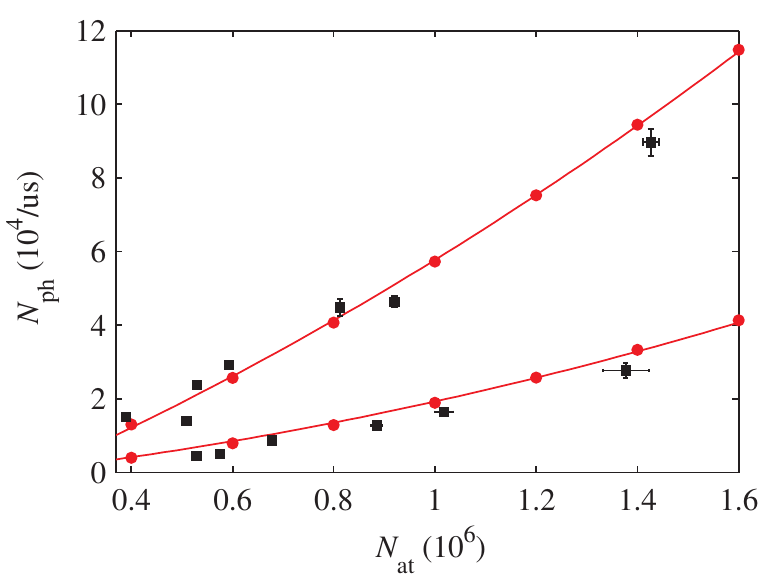}\\ 
  \caption{Atom number dependence of the peak value of the first superradiant pulse. (Black squares) Experimental data, (red circles) simulations, (red line) quadratic fit to simulations. For the lower data, $R =0.66  \times 10^3$ s$^{-1}$; for the upper data, \mbox{$R =2.2 \times 10^3$ s$^{-1}$.}  The representative errorbars show the standard error of the mean of three realizations for each setting.}
  \label{fig:atomnodep}
\end{figure}

Figure \ref{fig:atomnodep} shows the experimental and simulated amplitude of the first superradiant pulse as a function of atom number for two values of the single particle scattering rate. 
After a given hold time in the magnetic trap after condensation, three realizations of the unperturbed BEC followed by three realizations of superradiant scattering were performed. The atom number was obtained from fitting Thomas-Fermi profiles to the unperturbed expanded clouds and inferring the chemical potential \cite{Ketterle_review}. 
There is good overall agreement between the simulations and the data, and it is evident that the superradiant pulse amplitude  has the expected quadratic dependence in this parameter range.

We can see how  this quadratic dependence on the atom number arises by manipulating the 1D \MS equations. For simplicity, we consider the case where only the zeroth and first order atomic modes are populated.  This corresponds well to the low pump power case $R=0.45\times10^3$ s$^{-1}$ in Figure \ref{fig:data_forms}(a). Neglecting the wavefunction kinetic energy and displacement terms,  Equations (\ref{eqn:SR_wf})--(\ref{eqn:eminus}) in this parameter range  become:
\begin{align}
\frac{\partial\wfzeron}{\partial\tau} & = -i \Lambda\left[ \conjeplusn\eminusn\wftwon\rme^{-2i\tau} + (\abs{\eplusn}^2+\abs{\eminusn}^2)\wfzeron\right],\label{eqn:SR_wf_zero}\\
\frac{\partial\wftwon}{\partial\tau} & = -i \Lambda\left[ \conjeminusn\eplusn\wfzeron\rme^{+2i\tau}+
(\abs{\eplusn}^2+\abs{\eminusn}^2)\wftwon\right],\label{eqn:SR_wf_two}\\
\pdxi{\eplusn}& =  -i\frac{\Lambda}{\chi}\left[\eminusn\wftwon\conjwfzeron\rme^{-2i\tau}
+  (\abs{\wfzeron}^2+\abs{\wftwon}^2)\eplusn\right],\label{eqn:eplustwomode}\\
\pdxi{\eminusn}& =  +i\frac{\Lambda}{\chi} \left[\eplusn\wfzeron\conjwftwon\rme^{+2i\tau}
+  (\abs{\wfzeron}^2+\abs{\wftwon}^2)\eminusn\right].\label{eqn:eminustwomode}
\end{align}
Given that the growth of $e_-$ depends on the coherence (or polarization) term, $\psi_2\conj{\psi}_0$,  we consider:
\begin{equation}\label{eqn:pol_twomode}
 \frac{\partial(\wftwon\conjwfzeron)}{\partial\tau}
 = i \Lambda\conjeminusn\eplusn\rme^{+2i\tau}
 (\abs{\wftwon}^2-\abs{\wfzeron}^2).
\end{equation}
The first important feature is that the growth of the coherence term, which drives the creation of $e_-$ photons, depends on the population difference between the two momentum orders. In this way, we can regard  the condition for superradiant Rayleigh scattering as inversion in momentum space, as opposed to electronic state population inversion in Dicke superradiance. Evidently, the growth of the coherence $\psi_2\conj{\psi}_0$ is proportional to the number of atoms in the sample; since this term drives the growth of the field amplitude $e_-$, the backscattered photon flux is thus  proportional to $\Nat^2$ (see Equation~\eqref{eqn:Nph}).
We note further that the time development of the atomic coherence depends on the light field coherence $\conjeminusn\eplusn$. This is consistent with the physical picture that  both the local amplitude and phase of the matter and light wave coherences  determine the evolution of the system.

\section{Spatially dependent dynamics}\label{sect:spatial_dep_dynamics}
\begin{figure}[t]
\centering
  \includegraphics[width=15cm]{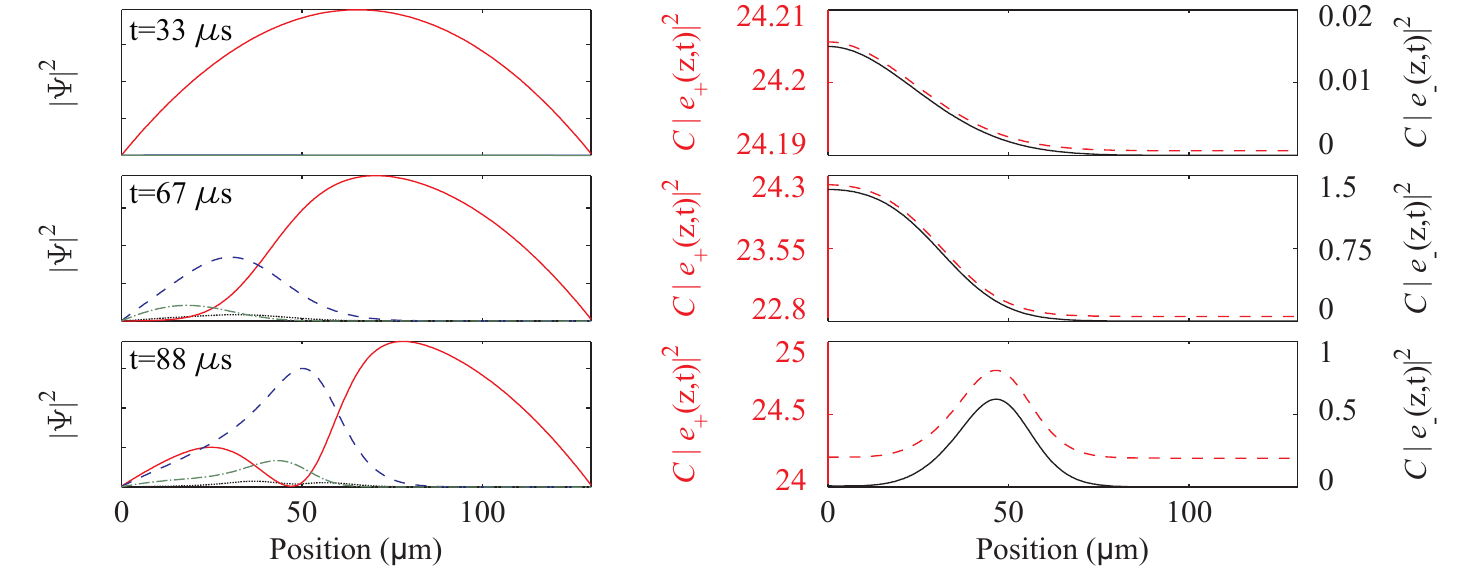}\\
  \caption{Simulated low power dynamics within the BEC at high detuning. The left column shows  the density of the different momentum components:  (black dotted line) $\psi_{-2}$, (red solid line) $\psi_{0}$ , (blue dashed line) $\psi_{2}$ ,  (green dash-dot line) $\psi_{+4}$. The right column shows the  light field intensity in units of $10^4/\mu$s: (red dashed line, left  $y$-axis)  $C\abs{e_+}^2$,  (black solid line, right $y$-axis) $C\abs{e_-}^2$. 
  The $x$-axis scale within the condensate (length $L=130 \mu$m) is the same for all the graphs.}
  \label{fig:sims_high_delta}
\end{figure}

To gain an understanding of the fundamental elements of the spatially dependent dynamics in the BEC, we present simulations for the low power case $R=0.45\times10^3$s$^{-1}$ where we experimentally observe that only the first order diffracted atomic mode is populated.
Figure \ref{fig:sims_high_delta} shows the results of our simulations. 
 For three relevant times during the interaction, atomic density distributions along the long axis of the BEC are shown in the left column, and the scaled intensities  $\propto\abs{e_+}^2$ and $\abs{e_-}^2$ are shown in the right column. (See also the Supplementary Online Material for a video of this simulation.) 
 
 The backscattered light intensity in the sample builds up at the  input end ($z=0$),  since  it sees gain from approximately the entire length of the BEC; this is the time shown in the top row  ($t=33~\mu$s). 
 This spatial inhomogeneity in the scattering can be directly observed in the atomic density distribution after modest  time-of-flight \cite{DominikSchneble04182003,zobay:041604,PhysRevA.73.013620,hilliard:051403}, because the expansion is slow along the symmetry axis of the condensate \cite{Castin:1996,Dalfovo:1997}.
 At this point, the rate of transfer of atoms from $\psi_0$ to $\psi_2$ concurrent with the growth of $e_-$ and reduction in $e_+$ begins to increase steeply. At the time shown in the second row ($t=67~\mu$s), the population in $\psi_0$ becomes sufficiently depleted at the input edge of the BEC that the process slows down and eventually stops. Thus, at this time, the first superradiant pulse reaches its peak amplitude and  begins thereafter to decay.
 However, the light field envelopes then move towards the centre of the condensate, where $\abs{\psi_0}$ is still large, and the exchange between the two light fields can  continue; this is the time shown in the bottom row ($t=88~\mu$s).  The result of this dynamics is that $\abs{\psi_0}$ grows again at the input end of the condensate,  driven there by the annihilation of  $\psi_2$ atoms and  $e_-$ photons that  were generated further inside the sample. This time marks the end of the first superradiant pulse. In the simulations,  superradiant ringing  arises from the repetition of these dynamics. 
 
 The light intensity in both $e_+$ and $e_-$ at $t=88~\mu$s shows a local maximum  within the BEC, implying that the dynamics leads to the formation of an optical resonator, where  partially reflecting mirrors are formed by the density modulation due to the interference of stationary and recoiling matter-waves. These Bragg gratings are centered where $\psi_0$ and $\psi_2$ cross.
However, this feature of the simulations is not manifest in the experimental data, as noted in Section \ref{sect:forms_data_sims}. Indeed, the experiments support the veracity of the simulations up to times shortly after the peak of the first superradiant pulse, but not the formation of a high-quality cavity by Bragg gratings. The 1D model neglects the transverse variation of these Bragg gratings, so that they resemble the paradigmatic Fabry-Perot etalon formed by two parallel flat mirrors.  In real cavities, the resonator stability   depends critically on the geometry of the mirrors  \cite{Las_phys}. We speculate that the disagreement  between simulations and experiment after the first superradiant pulse can be partially explained by the transverse variation of the Bragg gratings, in that they do not realize a  stable resonator. The slow decay of  light intensity after the peak of the superradiant pulse  in Figure \ref{fig:data_forms} is suggestive of decay in an optical cavity.
A full 3D model would allow for the inclusion of the transverse dynamics, effectively turning the single transverse mode description into a transverse multi-mode problem. This is an interesting but numerically intensive challenge, and beyond the scope of the present work.

\section{Conclusion and Outlook}\label{conclusion}

We have explored the semi-classical aspects of superradiant light scattering in  an end-pumped Bose Einstein condensate,
using time-resolved detection of the scattered light as a sensitive probe of the dynamics. The scaling of the  first superradiant pulse's amplitude and width  with the pump power was demonstrated, illustrating the analogous role the single particle scattering rate $R$ plays in SLS to the natural decay rate $\Gamma$ in Dicke superradiance. Additionally, we showed the quadratic scaling of the superradiant pulse amplitude with the sample atom number, which is characteristic of Dicke superradiance. The physical picture of four-wave mixing was developed using 1D \MS equations to model the dynamics. Despite their simplicity, the simulations achieve very good agreement with the experimentally observed superradiant pulse amplitude and timing. 
Experimentally, it would be interesting to detect the amplitude and phase (frequency) of the backscattered light via heterodyne detection to allow further comparison with the model where  the phase is an additional output parameter~\cite{PhysRevA.83.033615}.

In future work, we will explore the quantum aspects of the process.
Within the framework of the 1D model,  the quantum fluctuations that initiate the process may be included by seeding the dynamics using random initial conditions sampled from the appropriate probability distribution; in this case, the number distribution of photons of the thermal state of light that arises from spontaneous scattering. In the later dynamics, these random seeds manifest themselves in macroscopic fluctuations of arrival time and amplitude of the superradiant pulses~\cite{PhysRevA.20.2047,SR_review_GrossHaroche,PhysRevLett.83.5202,0953-4075-44-2-025301,PhysRevA.82.023608}.
From the microscopic quantum perspective,  backscattered photons and recoiling atoms are created in pairs much like the photon pairs in a two-mode squeezed state prepared by spontaneous down-conversion. This implies thermal statistics throughout the process for atoms and photons separately, as recently demonstrated in \cite{PhysRevA.90.013615} for atoms. Moreover, it implies  reduced, ideally vanishing, fluctuations for the number difference between the two modes. This has been shown for the closely related system of collisional four-wave mixing in \cite{PhysRevLett.105.190402}, but remains an open challenge for experiments with superradiant Bose Einstein condensates.

\section{Acknowledgements}\label{ack}
We thank Franziska Kaminski for her contribution to the experiment.
This work was supported by the DNRF centre QUANTOP and EU projects EMALI (MRTN-CT-2006-035369) and QAP.
J.A. acknowledges support from the Lundbeck Foundation.
D.W. acknowledges support from the Helmholtz Association via  grant no. VH-NG-1025.
\bibliographystyle{tMOP}

\end{document}